\documentclass[12pt,a4paper]{article}
\usepackage{amssymb}
\usepackage{amsmath}
\usepackage{latexsym}
\usepackage{cite}
\usepackage{physics}
\usepackage{float}
\usepackage{graphicx}
\begin{document}

\title{\vspace{-1cm}\bf Doubling of physical states in the quantum scalar field theory for a remote observer\\ in the Schwarzschild spacetime}

\author{
Vadim~Egorov, Mikhail~Smolyakov, Igor~Volobuev
\\
{\small{\em Skobeltsyn Institute of Nuclear Physics, Lomonosov Moscow
State University,
}}\\
{\small{\em Moscow 119991, Russia}}}

\date{}
\maketitle

\begin{abstract}
We discuss the problem of canonical quantization of a free real massive scalar field in the Schwarzschild spacetime. It is shown that a consistent procedure of canonical quantization of the field can be carried out without taking into account the black hole interior, so that in the resulting theory
the canonical commutation relations are satisfied exactly, and the Hamiltonian has the standard form. However, unlike some papers, in which the expansion of the quantum field in spherical harmonics is used, here we use an expansion in scatteringlike states for energies larger than the mass of the field. This reveals a strange property of the resulting quantum field theory --- doubling of the quantum states, which look as having the same asymptotic momentum to an observer located far away from the black hole. This purely topological effect cannot be eliminated by moving away from the black hole.
\end{abstract}

\section{Introduction}
The problem of quantization of fields in the presence of black
holes is widely discussed in the scientific literature, starting
from the pioneering papers \cite{Boulware:1974dm,HH}. However,
there still exist some unsolved problems within the standard
treatment of quantum theory in the black hole
background. For example, it is well known that in the
Kruskal–Szekeres coordinates \cite{Kruskal,Szekeres}, which
describe the maximal analytic extension of the Schwarzschild
spacetime, there exists a second, so called ``white
hole''. The latter reveals some problems with a physical
interpretation of the resulting theory, in particular, there
arises the well-known problem with locality, which is connected
with the location of the white hole in our universe
or even in a parallel world. The amount of scientific
literature on this topic is large, so we would like to highlight
the latest papers by G.~'t~Hooft \cite{tHooft1,tHooft2,tHooft3},
in which mathematically rigorous attempts to solve this problem
for the Schwarzschild solution were made. In papers
\cite{tHooft1,tHooft2} a geometrical identification of some areas
in the Kruskal–Szekeres spacetime (which was called ``antipodal
identification'') was proposed. Ideologically this identification
is very similar to orbifolding in models with extra
dimensions (see, for example, \cite{Rattazzi:2003ea}). However,
later it was shown \cite{tHooft3} that such an identification
leads to contradictions (namely, problems with
$CPT$--invariance). Instead of the antipodal identification, the
idea of ``quantum cloning'' of the black and white holes exteriors
was proposed in \cite{tHooft3}. An interesting property of the
approach is that the interior regions of both holes do not play
any role in the evolution and turn out to be mathematical
artifacts that do not have a direct physical
interpretation. However, the approach still has a drawback ---
there may emerge closed timelike curves \cite{tHooft3}.

According to the reasoning presented above (especially to the
observation that at least some rigorous approaches to obtaining a
consistent quantum theory in the black hole
background lead to the needlessness of the black and white hole
interiors), there arises a question about a
possibility to build a consistent quantum field theory in the
Schwarzschild spacetime outside the horizon only. As we will see
below, it is indeed possible even within the framework of
canonical quantization, which is quite a surprising result. At
least in the simplest case of a real massive scalar field, formally the
resulting theory turns out to be complete and self-consistent.

Unlike some recent papers
\cite{Akhmedov:2020ryq,Anempodistov:2020oki,Bazarov:2021rrb}, in
which the quantum scalar field is also considered
only outside the event horizon of the Schwarzschild black hole,
we use the field expansion in the scatteringlike states,
which are close to slightly modified plane waves, if we go far away from the black
hole. These states are very useful, because at large distances
from the black hole they are similar to the states that are used in
quantization of the scalar field in Minkowski
spacetime. This approach reveals an alarming feature of
the resulting quantum field theory --- doubling of the quantum
states, which look as having the same asymptotic momentum to a distant observer. It is a purely topological effect that
manifests itself even at such distances from the Schwarzschild
black hole at which one may naively expect that the
effects caused by the black hole can be neglected.

This paper is rather technical in the sense that it involves a
detailed examination of the spectrum of states outside the
Schwarzschild black hole, all canonical commutations relations are
checked to be exactly satisfied, and the resulting Hamiltonian is
obtained in the explicit form. All this is done in order to make
sure that the resulting theory is indeed self-consistent and does
not contain any contradictions.

\section{Setup}
Let us take a real massive scalar field $\phi(t,\vec x)$ in a
curved background described by the metric $g_{\mu\nu}$. The action
of the theory is
\begin{equation}
S=\int\mathcal{L}\,d^{4}x=\int\sqrt{-g}\left(\frac{1}{2}\,g^{\mu\nu}\partial_{\mu}\phi\,\partial_{\nu}\phi-\frac{M^{2}}{2}\phi^{2}\right)d^{4}x.
\end{equation}
Suppose that the metric $g_{\mu\nu}$ is static, i.e., it does not
depend on time. In such a case the equation of motion for the
scalar field takes the form
\begin{equation}\label{scalareqm}
\sqrt{-g}\,g^{00}\ddot\phi+\partial_{i}\left(\sqrt{-g}\,g^{ij}\partial_{j}\phi\right)+M^{2}\sqrt{-g}\,\phi=0,
\end{equation}
where $\dot\phi=\partial_{0}\phi$. It is clear that
the field $\phi(t,\vec x)$ can be expanded in solutions of the form
\begin{equation}
e^{\pm iE t}F(E,\vec x),
\end{equation}
where, without loss of generality, we can set $E\ge 0$. The latter
representation leads to the equation
\begin{equation}\label{scalareqm1}
-E^{2}\sqrt{-g}\,g^{00}F+\partial_{i}\left(\sqrt{-g}\,g^{ij}\partial_{j}F\right)+M^{2}\sqrt{-g}\,F=0,
\end{equation}
which, together with corresponding boundary
conditions, defines an eigenvalue problem. In particular,
Eq.~\eqref{scalareqm1} implies the following
orthogonality conditions for the solutions with
$E\neq E'$
\begin{equation}\label{orth0}
\int\sqrt{-g}\,g^{00}F^{*}(E,\vec x)F(E',\vec x)d^{3}x=0,\qquad \int\sqrt{-g}\,g^{00}F(E,\vec x)F(E',\vec x)d^{3}x=0,
\end{equation}
which can be easily obtained by multiplying the complex conjugate
of Eq.~\eqref{scalareqm1} (or Eq.~\eqref{scalareqm1} as it is) by
$F(E',\vec x)$, integrating the result with respect to $\vec x$ and
performing an integration by parts.

The component $T_{00}$ of the energy-momentum tensor
of a real massive scalar field has the standard form
\begin{equation}\label{T00scalar}
T_{00}=\frac{1}{2}\dot\phi^{2}-\frac{1}{2}g_{00}g^{ij}\partial_{i}\phi\,\partial_{j}\phi+\frac{M^{2}}{2}g_{00}\phi^{2}.
\end{equation}
It is well known that in General Relativity the covariant
conservation law is satisfied for any energy-momentum tensor
and can be rewritten as \cite{LL-FT}
\begin{equation}\label{tcons}
\nabla_{\mu}T^{\mu}_{\nu}=\frac{1}{\sqrt{-g}}\frac{\partial\left(\sqrt{-g}\,T^{\mu}_{\nu}\right)}{\partial x^{\mu}}-\frac{1}{2}\frac{\partial g_{\mu\sigma}}{\partial x^{\nu}}T^{\mu\sigma}=0.
\end{equation}
Since we are interested in the cases, in which the metric is
static (the Schwarzschild metric is exactly of this sort),
it follows from \eqref{tcons} that for $\nu=0$
\begin{equation}
\frac{\partial}{\partial x^{0}}\int\sqrt{-g}\,T^{0}_{0}d^{3}x=0.
\end{equation}
Thus, we can define the energy of the system, which is conserved over time (i.e., the Hamiltonian of the system), as
\begin{equation}\label{Hamilt}
H=\int\sqrt{-g}\,g^{00}T_{00}d^{3}x.
\end{equation}
Substituting the explicit expression \eqref{T00scalar} into \eqref{Hamilt}, performing an integration by parts and using equation of motion \eqref{scalareqm}, we arrive at
\begin{equation}\label{Hamiltscalar}
H=\frac{1}{2}\int\sqrt{-g}\,g^{00}\left(\dot\phi^{2}-\ddot\phi\,\phi\right)d^{3}x.
\end{equation}

\section{Solutions of the equation of motion}
\subsection{Properties of the spectrum}
Let us start with a detailed examination of the spectrum of
stationary states of Eq.~\eqref{scalareqm1}. Let us consider the
standard Schwarzschild metric and restrict ourselves to
the domain $r>r_{0}$, where $r_{0}$ is the
Schwarzschild radius. The field $F(E,\vec x)$ can be
expanded in spherical harmonics as
\begin{equation}\label{philm}
F(E,\vec x)=\sum\limits_{l=0}^{\infty}\sum\limits_{m=-l}^{l}\phi_{lm}^{}(E,r,\theta,\varphi)
=\sum\limits_{l=0}^{\infty}\sum\limits_{m=-l}^{l}Y_{lm}(\theta,\varphi)f_{l}(E,r),
\end{equation}
where (we use the convention of \cite{Korn-Korn})
\begin{equation}\label{Ylm}
Y_{lm}(\theta,\varphi)=\sqrt{\frac{2l+1}{4\pi}}\sqrt{\frac{(l-|m|)!}{(l+|m|)!}}\,P_{l}^{|m|}\left(\cos\theta\right)e^{im\varphi},\quad l=0,1,2, ... ,\quad m=0,\pm 1, \pm 2, ... ,
\end{equation}
leading to the radial equation
\begin{equation}\label{eqscalarrad}
E^{2}\frac{r}{r-r_{0}}f_{l}(E,r)-M^{2}f_{l}(E,r)+\frac{1}{r^{2}}\frac{d}{dr}\left(r(r-r_{0})\frac{df_{l}(E,r)}{dr}\right)
-\frac{l(l+1)}{r^{2}}f_{l}(E,r)=0
\end{equation}
for the functions $f_{l}(E,r)$. Without loss of generality we can
take $f_{l}(E,r)$ to be real. The orthogonality condition for
$f_{l}(E,r)$ is suggested by the form of Eq.~\eqref{eqscalarrad}
\begin{equation}
\int\limits_{r_{0}}^{\infty}\frac{r^{3}}{r-r_{0}}f_{l}(E,r)f_{l}(E',r)\,dr=0\quad\textrm{for}\quad E\neq E',
\end{equation}
as well as the form of the norm
\begin{equation}\label{norm}
\int\limits_{r_{0}}^{\infty}\frac{r^{3}}{r-r_{0}}f_{l}^{2}(E,r)\,dr.
\end{equation}

Using Eq.~\eqref{eqscalarrad}, it is possible to show that there
is no solution with $E=0$. Indeed, for $r\to r_{0}$ a possible
solution takes the form $f_{l}(0,r)\sim
1+\left(r_{0}M^{2}+\frac{l(l+1)}{r_{0}}\right)(r-r_{0})$, whereas
for $r\to\infty$ it takes the form
$f_{l}(0,r)\sim\frac{e^{-Mr}}{r}$. Let us multiply
Eq.~\eqref{eqscalarrad} by $r^{2}f_{l}(0,r)$,
integrate the result with respect to $r$ from $r_{0}$
to $\infty$, and perform an integration by parts. We get
\begin{equation}\label{intsurf}
r(r-r_{0})\frac{df_{l}(0,r)}{dr}f_{l}(0,r)\biggl|_{r_{0}}^{\infty}-\int\limits_{r_{0}}^{\infty}\left(r(r-r_{0})\left(\frac{df_{l}(0,r)}{dr}\right)^{2}
+\left(M^{2}r^{2}+l(l+1)\right)f_{l}^{2}(0,r)\right)dr=0.
\end{equation}
Since the surface terms in \eqref{intsurf} are equal to zero, we get
\begin{equation}\label{E0intrel}
\int\limits_{r_{0}}^{\infty}\left(r(r-r_{0})\left(\frac{df_{l}(0,r)}{dr}\right)^{2}
+\left(M^{2}r^{2}+l(l+1)\right)f_{l}^{2}(0,r)\right)dr=0.
\end{equation}
The integrand in \eqref{E0intrel} is non-negative for any $r$, which means that the only solution satisfying \eqref{E0intrel} is $f_{l}(0,r)\equiv 0$.

It should be noted that there is a controversy in the scientific
literature concerning the properties of the spectrum
and the eigenfunctions of radial equation
\eqref{eqscalarrad}. For example, in the well-known paper
\cite{Deruelle:1974zy} it is stated that the spectrum of states
for $E<M$ is discrete (though each state has an infinite norm). In
paper \cite{Zecca3} it is shown that the spectrum is continuous
and the radial solutions can be expressed in terms of the Heun
functions.\footnote{See also \cite{Fiziev:2005ki} for the Heun
functions in the case of Regge-Wheeler equation \cite{RW}.} In
paper \cite{GN} it is stated that from a quantum mechanical point
of view there exists the ``fall to the center'' regime
\cite{Case,Perelomov-Popov} on the event horizon making the whole
theory ill-behaved. To the best of our knowledge, the only paper
in which the properties of the spectrum of the radial equation
\eqref{eqscalarrad} are correctly described from a physical point
of view without going into details (i.e., without obtaining
explicit solutions like it was done in paper \cite{Zecca3}) is
\cite{Barranco:2011eyw}. So, below in this section we will
reproduce the results of \cite{Barranco:2011eyw}, though in more
detail.

\begin{figure}[ht]
\centering
\begin{minipage}{.49\textwidth}
\centering
\includegraphics[width=0.95\linewidth]{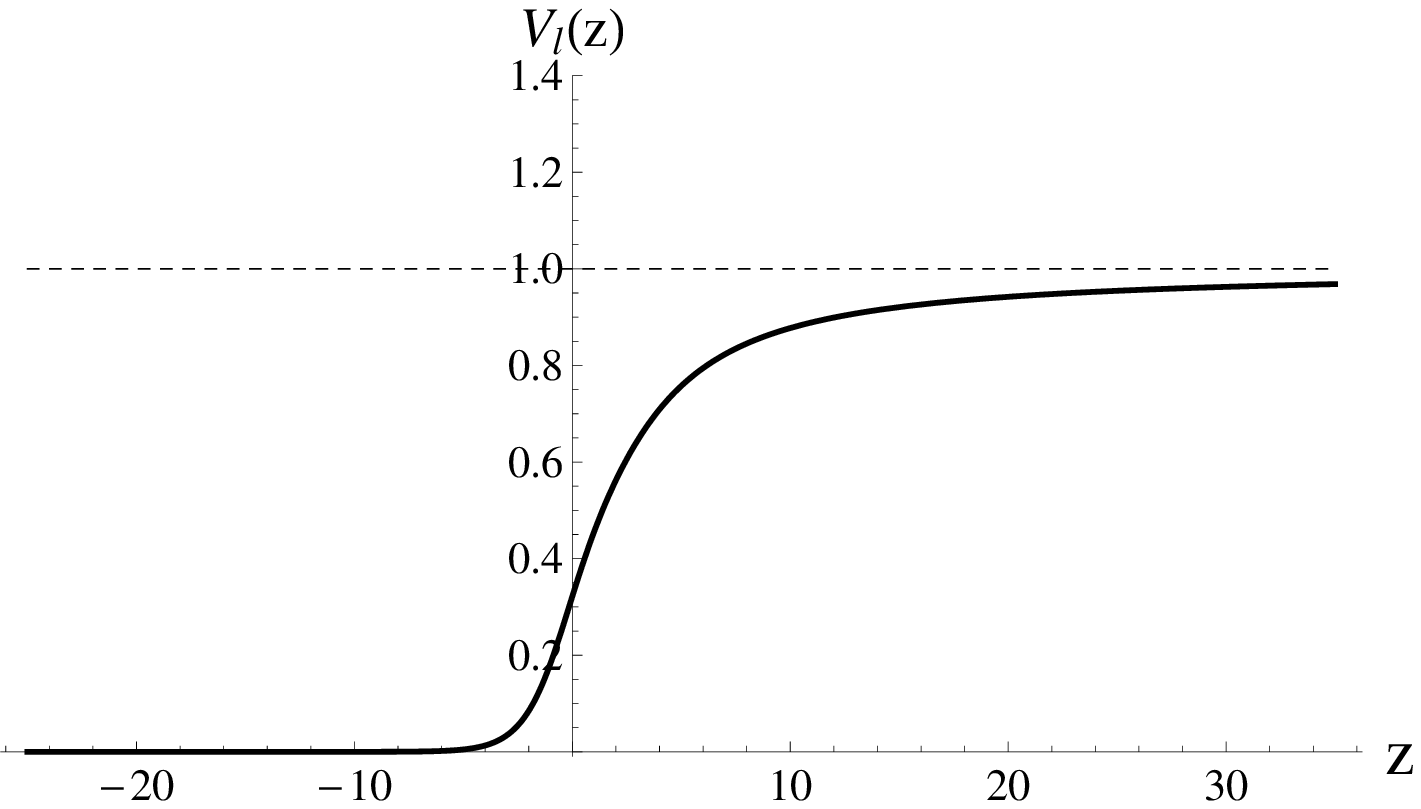}
\end{minipage}
\begin{minipage}{.49\textwidth}
\centering
\includegraphics[width=0.95\linewidth]{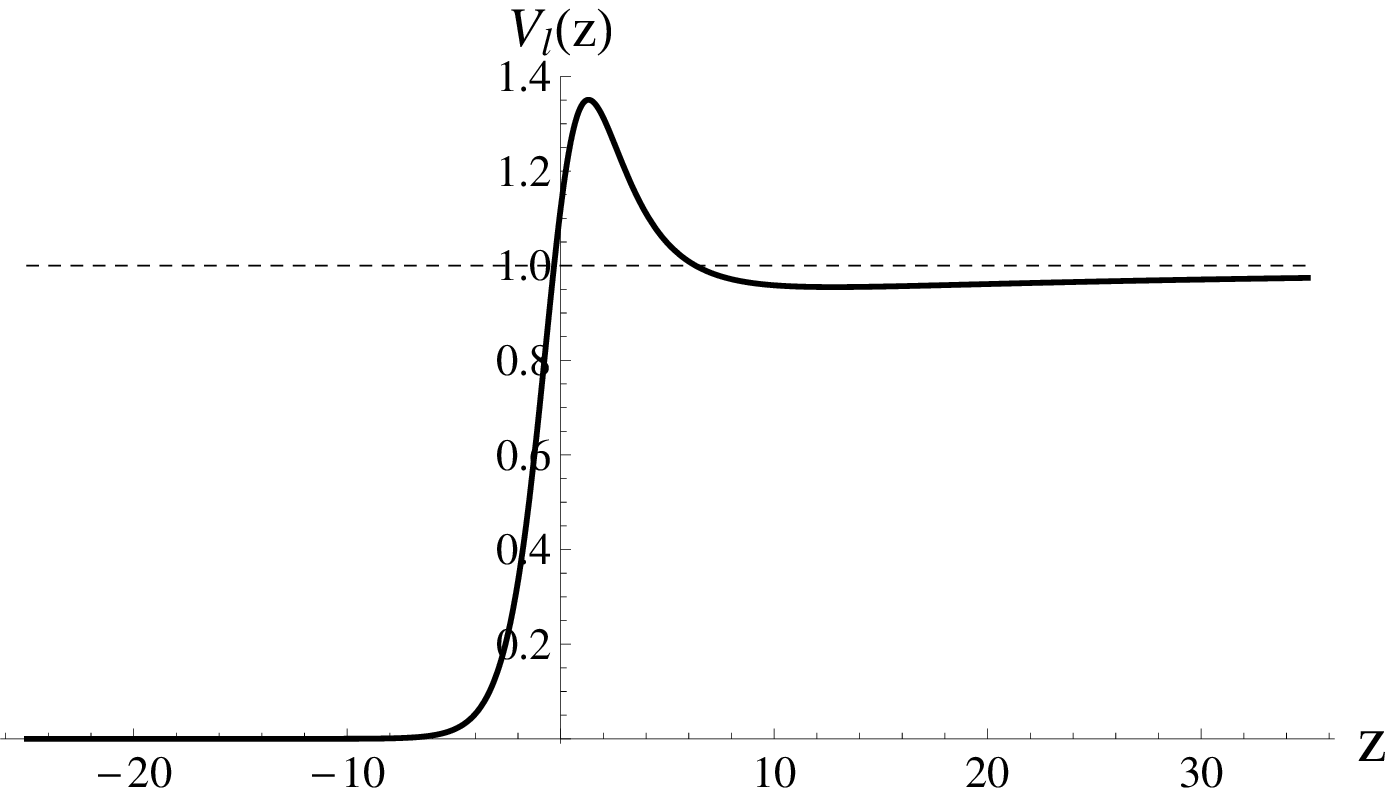}
\end{minipage}
\caption{$V_{l}(z)$ for $\mu=1$: $l=0$ (left plot) and $l=2$ (right plot). Dashed lines stand for $\mu^{2}$.}\label{fig2}
\end{figure}

First, let us introduce the dimensionless variables
\begin{equation}\label{substdimens}
\rho=\frac{r}{r_{0}},\qquad \mu=Mr_{0},\qquad \epsilon=E r_{0},\qquad u_{l}(\epsilon,\rho)=r_{0}f_{l}(E,r),
\end{equation}
where $\rho>1$. In these variables Eq.~\eqref{eqscalarrad} takes the form
\begin{equation}\label{eqscalarraddim}
-\frac{d}{d\rho}\left(\rho(\rho-1)\frac{du_{l}(\epsilon,\rho)}{d\rho}\right)+\left(\mu^{2}\rho^{2}+l(l+1)\right)u_{l}(\epsilon,\rho)
=\epsilon^{2}\frac{\rho^{3}}{\rho-1}u_{l}(\epsilon,\rho).
\end{equation}
Second, let us pass to the tortoise coordinate
\begin{equation}\label{turt}
z=\rho+\ln(\rho-1).
\end{equation}
Then Eq.~\eqref{eqscalarraddim} takes the form
\begin{equation}\label{eqShL}
-\frac{d}{dz}\left(\rho^{2}(z)\frac{du_{l}(\epsilon,z)}{dz}\right)+\frac{\rho(z)-1}{\rho(z)}\left(\mu^{2}\rho^{2}(z)+l(l+1)\right)u_{l}(\epsilon,z)=\epsilon^{2}\rho^{2}(z)u_{l}(\epsilon,z),
\end{equation}
where $\rho(z)$ is determined by \eqref{turt}. And third, using the substitution
\begin{equation}\label{substg}
u_{l}(\epsilon,z)=\frac{\psi_{l}(\epsilon,z)}{\rho(z)},
\end{equation}
Eq.~\eqref{eqShL} can be expressed in the form of a one-dimensional Schr\"{o}dinger equation
\begin{equation}\label{eqSchr}
-\frac{d^{2}\psi_{l}(\epsilon,z)}{dz^{2}}+V_{l}(z)\psi_{l}(\epsilon,z)=\epsilon^{2}\psi_{l}(\epsilon,z),
\end{equation}
where the potential has the form \cite{Barranco:2011eyw}
\begin{equation}\label{VSchr1}
V_{l}(z)=\frac{\rho(z)-1}{\rho(z)}\left(\mu^{2}+\frac{l(l+1)}{\rho^{2}(z)}+\frac{1}{\rho^{3}(z)}\right).
\end{equation}
The potential $V_{l}(z)$ is such that $V_{l}(z)\to 0$ for $z\to-\infty$ and $V_{l}(z)\to\mu^{2}$ for $z\to\infty$. In Fig.~\ref{fig2} some examples of $V_{l}(z)$ are presented. Fig.~\ref{fig2} also supports the fact that $f_{l}(0,r)\equiv 0$.
One can see that the asymptotic behavior of $V_{l}(z)$ corresponds
to a step-like potential, though there can be rises and dips in
the vicinity of $z=0$ depending on the value of $l$.\footnote{A
detailed discussion of scattering by the square
potential step can be found, for example, in \cite{Messiah}.} Such
potentials imply the continuity of the spectrum for $\epsilon>0$.

For the initial norm \eqref{norm} we get the result which is expected taking into account the form of Eq.~\eqref{eqSchr}:
\begin{equation}
\int\limits_{1}^{\infty}\frac{\rho^{3}}{\rho-1}\,u_{l}^{2}(\epsilon,\rho)d\rho=\int\limits_{-\infty}^{\infty}\rho^{2}(z)u_{l}^{2}(\epsilon,z)dz
=\int\limits_{-\infty}^{\infty}\psi_{l}^{2}(\epsilon,z)dz.
\end{equation}

Now let us turn to the discussion of the properties
of the radial solutions $\psi_{l}(\epsilon,z)$. For $\epsilon<\mu$
and $z\to\infty$ formally there exist two asymptotics $\sim
e^{\pm\sqrt{\mu^{2}-\epsilon^{2}}\,z}$ leading to a constant
Wronskian. However, the only solution which can be properly
normalized is the one with the asymptotics $\sim
e^{-\sqrt{\mu^{2}-\epsilon^{2}}\,z}$ at $z\to\infty$, for
$z\to-\infty$ the asymptotics of this solution is
$\sim\cos\left(\epsilon z-\gamma_{l}\right)$, where $\gamma_{l}$
is some phase. Thus, for a fixed $\epsilon$ and $l$ there exists
only one physically relevant solution. An important
point is that the spectrum of radial states for $0<\epsilon<\mu$
is continuous, which is provided by the phase $\gamma_{l}$
\cite{Barranco:2011eyw}.

For $\epsilon>\mu$ the situation is different. Since
the asymptotics of solutions are finite at
$z\to\pm\infty$, for fixed $\epsilon$ and $l$ there exist two
linearly independent solutions. In principle, these solutions can
be connected with solutions corresponding to the waves moving
towards $z\to\infty$ and towards $z\to-\infty$ (of course, one
should take into account one-dimensional scattering
by the potential $V_{l}(z)$ in these solutions)
\cite{Anempodistov:2020oki}. However, without loss of generality
these linearly independent solutions can be chosen to be real,
and we will denote them by $\psi_{lp}(\epsilon,z)$,
where $p=1,2$.

It is necessary to note that at $z\to-\infty$ (i.e., at $r\to r_{0}$) the potential $V_{l}(z)$ vanishes, which means that the field becomes effectively massless in this area.

\subsection{Orthogonality conditions and completeness relation for the\\ eigenfunctions}
It is clear that the functions $\psi_{l}(\epsilon,z)$ and $\psi_{lp}(\epsilon,z)$ can be normalized in such a way that the orthogonality conditions
\begin{align}\label{orthrad10psi}
\int\limits_{-\infty}^{\infty}\psi_{l}(\epsilon,z)\psi_{l}(\epsilon',z)dz&=\delta(\epsilon-\epsilon'),\\\label{orthrad20psi}
\int\limits_{-\infty}^{\infty}\psi_{l}(\epsilon,z)\psi_{lp}(\epsilon',z)dz&=0,\\\label{orthrad30psi}
\int\limits_{-\infty}^{\infty}\psi_{lp}(\epsilon,z)\psi_{lp'}(\epsilon',z)dz&=\delta_{pp'}\delta(\epsilon-\epsilon')
\end{align}
hold. Then, using \eqref{substdimens}, \eqref{turt}, and
\eqref{substg}, we get the orthogonality conditions
for the initial radial functions $f_{l}(E,r)$ and $f_{lp}(E,r)$
\begin{align}\label{orthrad10}
\int\limits_{r_{0}}^{\infty}f_{l}(E,r)f_{l}(E',r)\frac{r^{3}}{r-r_{0}}\,dr&=\delta(E-E'),\\\label{orthrad20}
\int\limits_{r_{0}}^{\infty}f_{l}(E,r)f_{lp}(E',r)\frac{r^{3}}{r-r_{0}}\,dr&=0,\\\label{orthrad30}
\int\limits_{r_{0}}^{\infty}f_{lp}(E,r)f_{lp'}(E',r)\frac{r^{3}}{r-r_{0}}\,dr&=\delta_{pp'}\delta(E-E').
\end{align}
With \eqref{philm}, we get the resulting orthogonality conditions
\begin{align}\label{orthrad1}
\int\limits_{0}^{2\pi}\int\limits_{0}^{\pi}\int\limits_{r_{0}}^{\infty}\phi_{lm}^{*}(E,r,\theta,\varphi)\phi_{l'm'}^{}(E',r,\theta,\varphi)\frac{r^{3}}{r-r_{0}}\sin\theta\,dr d\theta d\varphi&=\delta_{ll'}\delta_{mm'}\delta(E-E'),\\\label{orthrad2}
\int\limits_{0}^{2\pi}\int\limits_{0}^{\pi}\int\limits_{r_{0}}^{\infty}\phi_{lm}^{*}(E,r,\theta,\varphi)\phi_{l'm'p}^{}(E',r,\theta,\varphi)\frac{r^{3}}{r-r_{0}}\sin\theta\,dr d\theta d\varphi&=0,\\\label{orthrad3}
\int\limits_{0}^{2\pi}\int\limits_{0}^{\pi}\int\limits_{r_{0}}^{\infty}\phi_{lmp}^{*}(E,r,\theta,\varphi)\phi_{l'm'p'}^{}(E',r,\theta,\varphi)\frac{r^{3}}{r-r_{0}}\sin\theta\,dr d\theta d\varphi&=\delta_{pp'}\delta_{ll'}\delta_{mm'}\delta(E-E'),
\end{align}
where
\begin{equation}\label{phiYf}
\phi_{lmp}^{}(E,r,\theta,\varphi)=Y_{lm}(\theta,\varphi)f_{lp}(E,r).
\end{equation}

Since in Eq.~\eqref{eqSchr} we have a standard Hermitian
operator, the eigenfunctions of this equation form a complete set
(see, for example, \cite{Korn-Korn}). Taking into account the fact
that the normalization on the ``energy scale'' (not
the ``energy scale'' squared) was chosen in
\eqref{orthrad10psi}--\eqref{orthrad30psi}, the completeness
relation for the radial functions $\psi_{l}(\epsilon,z)$ also has
the standard form and looks like
\begin{equation}\label{complete1}
\int\limits_{0}^{\mu}\psi_{l}(\epsilon,z)\psi_{l}(\epsilon,z')\,d\epsilon+\sum\limits_{p=1}^{2}\int\limits_{\mu}^{\infty}\psi_{lp}(\epsilon,z)\psi_{lp}(\epsilon,z')\,d\epsilon=\delta(z-z').
\end{equation}
Consequently, the set of the corresponding solutions in the
initial coordinate $r$ also forms a complete set of
eigenfunctions, the corresponding completeness relation
\begin{equation}\label{complete2}
\int\limits_{0}^{M}f_{l}(E,r)f_{l}(E,r')\,dE
+\sum\limits_{p=1}^{2}\int\limits_{M}^{\infty}f_{lp}(E,r)f_{lp}(E,r')\,dE=\frac{r-r_{0}}{r^{3}}\,\delta(r-r')
\end{equation}
can be easily obtained by substituting \eqref{substdimens}, \eqref{turt}, and \eqref{substg} into \eqref{complete1}. Taking into account the angular parts of the eigenfunctions $Y_{lm}(\theta,\varphi)$, we can finally write
\begin{align}\nonumber
\sum\limits_{l=0}^{\infty}\sum\limits_{m=-l}^{l}&\left(\int\limits_{0}^{M}\phi_{lm}^{*}(E,r,\theta,\varphi)\phi_{lm}^{}(E,r',\theta',\varphi')\,dE
+\sum\limits_{p=1}^{2}\int\limits_{M}^{\infty}
\phi_{lmp}^{*}(E,r,\theta,\varphi)\phi_{lmp}^{}(E,r',\theta',\varphi')\,dE\right)\\
\label{complete3}
&=\frac{r-r_{0}}{r^{3}}\,
\delta(r-r')\delta(\cos\theta-\cos\theta')\delta(\varphi-\varphi').
\end{align}

\subsection{Scatteringlike states}
As we will see below, from a physical point of view, for
the energies larger than the mass of the field, it
is more convenient to use scatteringlike states instead of the
states $\phi_{lmp}^{}(E,r,\theta,\varphi)$ described above. Let us start with Eq.~\eqref{eqscalarrad}. At large $r$ Eq.~\eqref{eqscalarrad} can be rewritten as
\begin{equation}\label{eqscalarrad-large-r}
(E^{2}-M^{2})f_{lp}(E,r)+\frac{E^{2}r_{0}}{r}f_{lp}(E,r)+\frac{1}{r^{2}}\frac{d}{dr}\left(r(r-r_{0})\frac{df_{lp}(E,r)}{dr}\right)\approx 0,
\end{equation}
where we have retained only the leading terms in $r$. The term $\sim\frac{1}{r}$ cannot be neglected due to the long-range interaction provided by this term. For $E>M$, the leading solution to Eq.~\eqref{eqscalarrad-large-r} can be parametrized as
\begin{equation}\label{sol-large-r}
f_{lp}(E,r)\approx C_{lp}^{+}(k)\frac{1}{r}\sin\left(kr+\frac{(2k^{2}+M^{2})r_{0}}{2k}\ln(kr)-\frac{\pi l}{2}+\tilde\delta_{lp}(k)\right),
\end{equation}
where $C_{lp}^{+}(k)$ are normalization constants, $k=\sqrt{E^{2}-M^{2}}$ and $\tilde\delta_{lp}(k)$ are phase shifts. Solution \eqref{sol-large-r} is analogous to the one for the standard Coulomb potential in quantum mechanics \cite{LL-QM}.

For $r\to r_{0}$ Eq.~\eqref{eqscalarrad} can be rewritten as
\begin{equation}\label{eqscalarrad-rr0}
\frac{E^{2}r_{0}^{2}}{r-r_{0}}f_{l}(E,r)+\frac{d}{dr}\left((r-r_{0})\frac{df_{l}(E,r)}{dr}\right)\approx 0,
\end{equation}
and its solution can be parametrized as
\begin{equation}\label{sol-rr0}
f_{lp}(E,r)\approx C_{lp}^{-}(k)\sin\left(\sqrt{k^{2}+M^{2}}\,r_{0}\ln(k(r-r_{0}))+\gamma_{lp}(k)\right),
\end{equation}
where $C_{lp}^{-}(k)$ are normalization constants and $\gamma_{lp}(k)$ are phase shifts.

It is clear that the normalization constants $C_{lp}^{+}(k)$ and $C_{lp}^{-}(k)$ are determined by the normalization integral at $r\to\infty$ and $r\to r_{0}$. The structure of approximate solutions \eqref{sol-large-r} and \eqref{sol-rr0} suggests that the phases can be chosen so that the normalization constants $C_{lp}^{+}(k)$ and $C_{lp}^{-}(k)$ do not depend on $l$, so we can write $C_{p}^{+}(k)$ and $C_{p}^{-}(k)$ for all $l$.

Now we turn to the scatteringlike states. Let us define these states exactly as in the standard scattering theory \cite{LL-QM},
\begin{equation}\label{scatstatesdec}
\phi_{p}(\vec k,\vec x)=\frac{1}{4\pi k}\sum\limits_{l=0}^{\infty}(2l+1)e^{i\left(\frac{\pi l}{2}+\tilde\delta_{lp}(k)\right)}P_{l}\left(\frac{\vec k\vec x}{kr}\right)\left(\frac{\sqrt{k}}{\left(k^2+M^{2}\right)^{\frac{1}{4}}}\,f_{lp}\left(\sqrt{k^{2}+M^{2}},r\right)\right),
\end{equation}
where $P_{l}(...)$ are the Legendre polynomials, $\tilde\delta_{lp}(k)$ are phase shifts defined by representation \eqref{sol-large-r}, $k=|\vec k|$, $r=|\vec x|$, and $\vec n=\frac{\vec x}{r}$. The extra factor $\frac{\sqrt{k}}{\left(k^2+M^{2}\right)^{1/4}}$ in \eqref{scatstatesdec} is introduced in order to have
\begin{equation}
\int\limits_{r_{0}}^{\infty}\left(\frac{\sqrt{k}\,f_{lp}\left(\sqrt{k^{2}+M^{2}},r\right)}{\left(k^2+M^{2}\right)^{\frac{1}{4}}}\right)
\left(\frac{\sqrt{k'}\,f_{lp'}\left(\sqrt{{k'}^{2}+M^{2}},r\right)}{\left({k'}^2+M^{2}\right)^{\frac{1}{4}}}\right)\frac{r^{3}}{r-r_{0}}\,dr=\delta_{pp'}\delta(k-k')
\end{equation}
and, consequently, to get a more physically reasonable normalization for the scatteringlike states. It is evident that $\phi_{p}(\vec k,\vec x)$ defined by \eqref{scatstatesdec} is a solution of Eq.~\eqref{scalareqm1} with $E=\sqrt{k^{2}+M^{2}}$. At large $r$
\begin{equation}\label{scatstatesdec1}
\phi_{p}(\vec k,\vec x)\sim\frac{1}{kr}\sum\limits_{l=0}^{\infty}(2l+1)e^{i\left(\frac{\pi l}{2}+\tilde\delta_{lp}(k)\right)}P_{l}\left(\frac{\vec k\vec x}{kr}\right)\sin\left(kr+\frac{(2k^{2}+M^{2})r_{0}}{2k}\ln(kr)-\frac{\pi l}{2}+\tilde\delta_{lp}(k)\right),
\end{equation}
where we have used approximate solution \eqref{sol-large-r}. Using the fact that at large $r$ (see, for example, \cite{LL-QM})
\begin{equation}\label{expdeclarge-r}
e^{i\vec k\vec x}\approx\frac{1}{kr}\sum\limits_{l=0}^{\infty}(2l+1)e^{\frac{i\pi l}{2}}P_{l}\left(\frac{\vec k\vec x}{kr}\right)\sin\left(kr-\frac{\pi l}{2}\right),
\end{equation}
one can easily show that
\begin{equation}\label{scatstates}
\phi_{p}(\vec k,\vec x)\sim e^{i\left(\vec k\vec x-\frac{(2k^{2}+M^{2})r_{0}}{2k}\ln(kr)\right)}+A_{p}(\vec k,\vec n,r)\frac{e^{ikr}}{r},\qquad p=1,2,
\end{equation}
where the functions $A_{p}(\vec k,\vec n,r)$ are defined as
\begin{equation}\label{scatampl}
A_{p}(\vec k,\vec n,r)=\frac{1}{2ik}\sum\limits_{l=0}^{\infty}(2l+1)P_{l}\left(\frac{\vec k\vec x}{kr}\right)\left(
e^{i\left(2\tilde\delta_{lp}(k)+\frac{(2k^{2}+M^{2})r_{0}}{2k}\ln(kr)\right)}-e^{-i\frac{(2k^{2}+M^{2})r_{0}}{2k}\ln(kr)}\right).
\end{equation}
These functions look similar to the standard scattering amplitudes, but they explicitly depend on $r$ (through the terms with $\ln(kr)$) and thus can not be considered as actual scattering amplitudes. The fact that we have the modified plane wave $e^{i\left(\vec k\vec x-\frac{(2k^{2}+M^{2})r_{0}}{2k}\ln(kr)\right)}$ instead of a simple plane wave in \eqref{scatstates} reflects the influence of the long-range potential $\sim\frac{1}{r}$ in \eqref{eqscalarrad-large-r}, which modifies the plane wave even at large distances from the black hole similarly to the case of the standard Coulomb potential in quantum mechanics \cite{LL-QM}.\footnote{This modified plane wave in the leading order is a solution of the initial equation of motion for the scalar field for $r\to\infty$ (which is just the Klein-Gordon equation). This also supports the assertion that the radial solutions can be chosen so that their normalization constants do not depend on $l$.}

We would like to stress that, since for fixed $l$ and $k$ there
exist two different solutions
$f_{l1}\left(\sqrt{k^{2}+M^{2}},r\right)$ and
$f_{l2}\left(\sqrt{k^{2}+M^{2}},r\right)$, for a fixed $\vec k$ we
can build {\em two} scatteringlike states of form \eqref{scatstates}
which differ in the functions $A_{p}(\vec k,\vec n,r)$.

It is clear that
\begin{equation}\label{orthscatt1}
\int\limits_{r>r_{0}}\sqrt{-g}\,g^{00}\phi_{lm}^{*}(E,\vec x)\phi_{p}^{}(\vec k,\vec x)\,d^{3}x=0,
\end{equation}
because $E$ in $\phi_{lm}(E,\vec x)$ and $\sqrt{k^{2}+M^{2}}$
belong to different energy ranges ($\sqrt{k^{2}+M^{2}}>M$ and $E<M$). Thus, the orthogonality condition
\eqref{orthscatt1} follows directly from \eqref{orth0}. In
\eqref{orthscatt1} the notation
\begin{equation}
\int\limits_{r>r_{0}}\sqrt{-g}\,g^{00}d^{3}x=\int\limits_{r_{0}}^{\infty}\frac{r^{3}}{r-r_{0}}\,dr
\int\limits_{0}^{\pi}\sin\theta d\theta
\int\limits_{0}^{2\pi}d\varphi
\end{equation}
is used. The scatteringlike states \eqref{scatstatesdec} are also orthogonal, the corresponding orthogonality condition takes the form
\begin{equation}\label{orthscatt2}
\int\limits_{r>r_{0}}\sqrt{-g}\,g^{00}\phi_{p}^{*}(\vec k,\vec x)\phi_{p'}^{}(\vec k',\vec x)\,d^{3}x=\delta_{pp'}\delta^{(3)}(\vec k-\vec k'),
\end{equation}
a detailed proof can be found in Appendix~A. The completeness
relation involving the scatteringlike states has the
form
\begin{equation}\label{complete4}
\sum\limits_{l=0}^{\infty}\sum\limits_{m=-l}^{l}\int\limits_{0}^{M}\phi_{lm}^{*}(E,\vec x)\phi_{lm}^{}(E,\vec y)\,dE+\sum\limits_{p=1}^{2}\int\phi_{p}^{*}(\vec k,\vec x)\phi_{p}^{}(\vec k,\vec y)\,d^{3}k=\frac{\delta^{(3)}(\vec x-\vec y)}{\sqrt{-g}\,g^{00}},
\end{equation}
a detailed proof can be found in Appendix~B. In \eqref{complete4} the notation
\begin{equation}
\int d^{3}k=\int\limits_{-\infty}^{\infty}dk_{1}\int\limits_{-\infty}^{\infty}dk_{2}\int\limits_{-\infty}^{\infty}dk_{3}=\int\limits_{0}^{\infty}k^{2}dk
\int\limits_{0}^{\pi}\sin\theta_{k}d\theta_{k}
\int\limits_{0}^{2\pi}d\varphi_{k},
\end{equation}
where $\theta_{k}$ and $\varphi_{k}$ are the angles in spherical coordinates in the momentum space, is used. The completeness relation \eqref{complete4} is necessary for performing a consistent procedure of quantization.

\section{Canonical quantization}
\subsection{Expansion of the quantum field}
Usually, when a quantum theory outside the horizon of the
Schwarzschild black hole is considered, the quantum scalar field
$\phi(t,\vec x)$ is expanded in spherical harmonics,
i.e., an expansion of the form
\begin{align}\nonumber
\phi(t,r,\theta,\varphi)=\sum\limits_{l=0}^{\infty}\sum\limits_{m=-l}^{l}\int\limits_{0}^{M}\frac{dE}{\sqrt{2E}}\left(e^{-iEt}\phi_{lm}^{}(E,r,\theta,\varphi)a_{lm}^{}(E)
+e^{iEt}\phi_{lm}^{*}(E,r,\theta,\varphi)a_{lm}^{\dagger}(E)\right)&\\\label{operatordecspher}
+\sum\limits_{p=1}^{2}\sum\limits_{l=0}^{\infty}\sum\limits_{m=-l}^{l}\int\limits_{M}^{\infty}\frac{dE}{\sqrt{2E}}\left(e^{-iEt}
\phi_{lmp}^{}(E,r,\theta,\varphi)a_{lmp}^{}(E)+e^{iEt}\phi_{lmp}^{*}(E,r,\theta,\varphi)a_{lmp}^{\dagger}(E)\right)&,
\end{align}
is used, see
\cite{Akhmedov:2020ryq,Anempodistov:2020oki,Bazarov:2021rrb}.\footnote{The same when the black hole interior is taken into account, see, for example, \cite{Giddings:2022sss}.} This
expansion is quite natural for a spherically
symmetric system like the one described by the Schwarzschild
background. However, this expansion is not useful
for examining the theory far away from the black hole, where we
expect that the influence of the black hole can be neglected.
Indeed, in Minkowski spacetime we prefer to use the
expansion in terms of plane waves, which provides a much more
convenient description of quantum states. Of course, because of
the Schwarzschild background we cannot use exactly the plane
waves, however, we can use the scatteringlike states
described in the previous section for the expansion. These
states behave like slightly modified plane waves far away from what is considered as
a ``central potential'', thus allowing one to describe particles
in the usual manner in that area but rigorously take into account
the effect produced by the ``potential''. Namely, let us take
\begin{align}\nonumber
\phi(t,\vec x)=\sum\limits_{l=0}^{\infty}\sum\limits_{m=-l}^{l}\int\limits_{0}^{M}\frac{dE}{\sqrt{2E}}\left(e^{-iEt}\phi_{lm}^{}(E,\vec x)a_{lm}^{}(E)+e^{iEt}\phi_{lm}^{*}(E,\vec x)a_{lm}^{\dagger}(E)\right)&\\\label{operatordec}
+\sum\limits_{p=1}^{2}\int\frac{d^{3}k}{\sqrt{2\sqrt{k^{2}+M^{2}}}}\left(e^{-i\sqrt{k^{2}+M^{2}}\,t}\phi_{p}^{}(\vec k,\vec x)a_{p}^{}(\vec k)+
e^{i\sqrt{k^{2}+M^{2}}\,t}\phi_{p}^{*}(\vec k,\vec x)a_{p}^{\dagger}(\vec k)\right)&,
\end{align}
where $\phi_{lm}^{}(E,\vec x)=\phi_{lm}^{}(E,r,\theta,\varphi)$ is
defined by \eqref{philm} with \eqref{Ylm}, $\phi_{p}^{}(\vec
k,\vec x)$ is defined by \eqref{scatstatesdec}. We suppose that
the creation and annihilation operators satisfy the
standard commutation relations
\begin{align}\label{CRa1}
&[a_{lm}^{}(E),a_{l'm'}^{\dagger}(E')]=\delta_{ll'}\delta_{mm'}\delta(E-E'),\\\label{CRa2}
&[a_{p}^{}(\vec k),a_{p'}^{\dagger}({\vec k}')]=\delta_{pp'}\delta^{(3)}(\vec k-\vec k'),
\end{align}
all other commutators being equal to zero. As we will see below,
this expansion is indeed more useful from a physical
point of view.

\subsection{Canonical commutation relations}
A consistent procedure of canonical quantization demands that the
canonical commutation relations are exactly satisfied. Let us
check that it is indeed so for expansion
\eqref{operatordec}. The canonical coordinate in a scalar field
theory is $\phi(t,\vec x)$, whereas the canonically conjugate
momentum is
\begin{equation}\label{ccm}
\pi(t,\vec x)\equiv\frac{\partial\mathcal{L}}{\partial\dot\phi(t,\vec x)}=\sqrt{-g(\vec x)}\,g^{00}(\vec x)\dot\phi(t,\vec x).
\end{equation}
The following canonical commutation relations should be satisfied:
\begin{equation}\label{CCR}
[\phi(t,\vec x),\pi(t,\vec y)]=i\delta^{(3)}(\vec x-\vec y),\qquad
[\phi(t,\vec x),\phi(t,\vec y)]=0,\qquad
[\pi(t,\vec x),\pi(t,\vec y)]=0.
\end{equation}
Substituting \eqref{ccm} into \eqref{CCR}, we get
\begin{align}\label{CCR1}
&[\phi(t,\vec x),\dot\phi(t,\vec y)]=i\frac{\delta^{(3)}(\vec x-\vec y)}{\sqrt{-g}\,g^{00}},\\\label{CCR2}
&[\phi(t,\vec x),\phi(t,\vec y)]=0,\\\label{CCR3}
&[\dot\phi(t,\vec x),\dot\phi(t,\vec y)]=0.
\end{align}

Let us check that commutation relations \eqref{CCR1}--\eqref{CCR3} are satisfied for \eqref{operatordec}. Substituting \eqref{operatordec} into the lhs of \eqref{CCR1} and using \eqref{CRa1} and \eqref{CRa2}, one gets
\begin{align}\nonumber
[\phi(t,\vec x),\dot\phi(t,\vec y)]=\frac{1}{2}\left(
\sum\limits_{l=0}^{\infty}\sum\limits_{m=-l}^{l}\int\limits_{0}^{M}\Bigl(\phi_{lm}^{}(E,\vec x)\phi_{lm}^{*}(E,\vec y)+\phi_{lm}^{}(E,\vec y)\phi_{lm}^{*}(E,\vec x)\Bigr)dE\right.\\\label{CCR1calc}\left.+
\sum\limits_{p=1}^{2}\int\Bigl(\phi_{p}^{}(\vec k,\vec x)\phi_{p}^{*}(\vec k,\vec y)+\phi_{p}^{}(\vec k,\vec y)\phi_{p}^{*}(\vec k,\vec x)\Bigr)d^{3}k
\right).
\end{align}
With the completeness relation \eqref{complete4}, for \eqref{CCR1calc} we get exactly \eqref{CCR1}.

Substituting \eqref{operatordec} into the lhs of \eqref{CCR2} and \eqref{CCR3}, and using \eqref{CRa1} and \eqref{CRa2}, one gets
\begin{align}\nonumber
[\phi(t,\vec x),\phi(t,\vec y)]=
\frac{1}{2}\sum\limits_{l=0}^{\infty}\sum\limits_{m=-l}^{l}\int\limits_{0}^{M}\Bigl(\phi_{lm}^{}(E,\vec x)\phi_{lm}^{*}(E,\vec y)-\phi_{lm}^{}(E,\vec y)\phi_{lm}^{*}(E,\vec x)\Bigr)\frac{dE}{E}&\\\label{CCR2calc}+
\frac{1}{2}\sum\limits_{p=1}^{2}\int\Bigl(\phi_{p}^{}(\vec k,\vec x)\phi_{p}^{*}(\vec k,\vec y)-\phi_{p}^{}(\vec k,\vec y)\phi_{p}^{*}(\vec k,\vec x)\Bigr)\frac{d^{3}k}{\sqrt{k^{2}+M^{2}}}&,
\end{align}
\begin{align}\nonumber
[\dot\phi(t,\vec x),\dot\phi(t,\vec y)]=
\frac{1}{2}\sum\limits_{l=0}^{\infty}\sum\limits_{m=-l}^{l}\int\limits_{0}^{M}\Bigl(\phi_{lm}^{}(E,\vec x)\phi_{lm}^{*}(E,\vec y)-\phi_{lm}^{}(E,\vec y)\phi_{lm}^{*}(E,\vec x)\Bigr)E\,dE&\\\label{CCR3calc}+
\frac{1}{2}\sum\limits_{p=1}^{2}\int\Bigl(\phi_{p}^{}(\vec k,\vec x)\phi_{p}^{*}(\vec k,\vec y)-\phi_{p}^{}(\vec k,\vec y)\phi_{p}^{*}(\vec k,\vec x)\Bigr)\sqrt{k^{2}+M^{2}}\,d^{3}k&.
\end{align}
Using formula \eqref{complete4app6} from Appendix~B (see also \eqref{complete4app7}), the double sum in \eqref{CCR2calc} can be represented as
\begin{align}\nonumber
\sum\limits_{l=0}^{\infty}\sum\limits_{m=-l}^{l}\int\limits_{0}^{M}\Bigl(\phi_{lm}^{}(E,\vec x)\phi_{lm}^{*}(E,\vec y)-\phi_{lm}^{}(E,\vec y)\phi_{lm}^{*}(E,\vec x)\Bigr)\frac{dE}{E}&\\\label{CCRcalcA}=\frac{1}{4\pi}\sum\limits_{l=0}^{\infty}(2l+1)P_{l}(\cos\gamma)
\int\limits_{0}^{M}\Bigl(f_{l}(E,r)f_{l}(E,r')-f_{l}(E,r')f_{l}(E,r)\Bigr)\frac{dE}{E}&=0.
\end{align}
Analogously, the double sum in \eqref{CCR3calc} can be represented as
\begin{align}\nonumber
\sum\limits_{l=0}^{\infty}\sum\limits_{m=-l}^{l}\int\limits_{0}^{M}\Bigl(\phi_{lm}^{}(E,\vec x)\phi_{lm}^{*}(E,\vec y)-\phi_{lm}^{}(E,\vec y)\phi_{lm}^{*}(E,\vec x)\Bigr)E\,dE&\\=\frac{1}{4\pi}\sum\limits_{l=0}^{\infty}(2l+1)P_{l}(\cos\gamma)
\int\limits_{0}^{M}\Bigl(f_{l}(E,r)f_{l}(E,r')-f_{l}(E,r')f_{l}(E,r)\Bigr)E\,dE&=0.
\end{align}
Next, using formulas \eqref{complete4app2}--\eqref{complete4app3} from Appendix~B, the second line in \eqref{CCR2calc} can be represented as
\begin{align}\nonumber
\sum\limits_{p=1}^{2}\int\Bigl(\phi_{p}^{}(\vec k,\vec x)\phi_{p}^{*}(\vec k,\vec y)-\phi_{p}^{}(\vec k,\vec y)\phi_{p}^{*}(\vec k,\vec x)\Bigr)\frac{d^{3}k}{\sqrt{k^{2}+M^{2}}}&\\=
\frac{1}{4\pi}\sum\limits_{l=0}^{\infty}(2l+1)
P_{l}(\cos\gamma)\sum\limits_{p=1}^{2}\int\limits_{M}^{\infty}\Bigl(
f_{lp}\left(E,r\right)f_{lp}\left(E,r'\right)-f_{lp}\left(E,r'\right)f_{lp}\left(E,r\right)\Bigr)\frac{dE}{E}&=0,
\end{align}
whereas the second line in \eqref{CCR3calc} can be represented as
\begin{align}\nonumber
\sum\limits_{p=1}^{2}\int\Bigl(\phi_{p}^{}(\vec k,\vec x)\phi_{p}^{*}(\vec k,\vec y)-\phi_{p}^{}(\vec k,\vec y)\phi_{p}^{*}(\vec k,\vec x)\Bigr)\sqrt{k^{2}+M^{2}}\,d^{3}k&\\\label{CCRcalcD}=
\frac{1}{4\pi}\sum\limits_{l=0}^{\infty}(2l+1)
P_{l}(\cos\gamma)\sum\limits_{p=1}^{2}\int\limits_{M}^{\infty}\Bigl(
f_{lp}\left(E,r\right)f_{lp}\left(E,r'\right)-f_{lp}\left(E,r'\right)f_{lp}\left(E,r\right)\Bigr)E\,dE&=0.
\end{align}
Substituting \eqref{CCRcalcA}--\eqref{CCRcalcD} into \eqref{CCR2calc} and \eqref{CCR3calc}, we get exactly \eqref{CCR2} and \eqref{CCR3}. Thus, all three canonical commutation relations are exactly satisfied for \eqref{operatordec}.

\subsection{Hamiltonian}
Now we turn to the calculation of the Hamiltonian of the system. Note that in addition to the orthogonality condition \eqref{orthscatt1}, the orthogonality conditions
\begin{equation}\label{orthcondaux}
\int\limits_{r>r_{0}}\sqrt{-g}g^{00}\phi_{lm}^{}(E,\vec x)\phi_{p}^{}(\vec k,\vec x)\,d^{3}x=0,\qquad \int\limits_{r>r_{0}}\sqrt{-g}g^{00}\phi_{lm}^{*}(E,\vec x)\phi_{p}^{*}(\vec k,\vec x)\,d^{3}x=0
\end{equation}
hold as well. They are also a consequence of the
fact that $E$ in $\phi_{lm}(E,\vec x)$ and $\sqrt{k^{2}+M^{2}}$
belong to different energy ranges ($\sqrt{k^{2}+M^{2}}>M$ and $E<M$) and follow directly from \eqref{orth0}.
Substituting \eqref{operatordec} into \eqref{Hamiltscalar} and
using the orthogonality conditions \eqref{orthrad1},
\eqref{orthscatt1}, \eqref{orthscatt2} and \eqref{orthcondaux},
after straightforward calculations one gets
\begingroup
\allowdisplaybreaks
\begin{align}\nonumber
H&=\frac{1}{2}\sum\limits_{l=0}^{\infty}\sum\limits_{m=-l}^{l}\int\limits_{0}^{M}E\left(a_{lm}^{\dagger}(E)a_{lm}^{}(E)+a_{lm}^{}(E)a_{lm}^{\dagger}(E)\right)dE
\\\nonumber
&+\frac{1}{2}\sum\limits_{p=1}^{2}\int\sqrt{k^{2}+M^{2}}
\left(a_{p}^{\dagger}(\vec k)a_{p}^{}(\vec k)+a_{p}^{}(\vec k)a_{p}^{\dagger}(\vec k)\right)d^{3}k
\\\nonumber
&+\frac{1}{4}\sum\limits_{l=0}^{\infty}\sum\limits_{m=-l}^{l}\sum\limits_{l'=0}^{\infty}\sum\limits_{m'=-l'}^{l'}\int\limits_{0}^{M}dE\int\limits_{0}^{M}dE'
\left(\frac{E\sqrt{E}}{\sqrt{E'}}-\sqrt{EE'}\right)
\\\nonumber
&\times\left(e^{-i(E+E')t}a_{lm}^{}(E)a_{l'm'}^{}(E')\int\limits_{r>r_{0}}\sqrt{-g}g^{00}\phi_{lm}^{}(E,\vec x)\phi_{l'm'}^{}(E',\vec x)\,d^{3}x+\textrm{H.c.}\right)\\\nonumber
&+\frac{1}{4}\sum\limits_{p=1}^{2}\sum\limits_{p'=1}^{2}\int d^{3}k\int d^{3}k'\left(\frac{(k^{2}+M^{2})^{\frac{3}{4}}}{({k'}^{2}+M^{2})^{\frac{1}{4}}}
-(k^{2}+M^{2})^{\frac{1}{4}}({k'}^{2}+M^{2})^{\frac{1}{4}}\right)\\\label{Hamaux1}
&\times\left(e^{-i\left(\sqrt{k^{2}+M^{2}}+\sqrt{{k'}^{2}+M^{2}}\right)t}a_{p}^{}(\vec k)a_{p'}^{}(\vec k')\int\limits_{r>r_{0}}\sqrt{-g}g^{00}
\phi_{p}^{}(\vec k,\vec x)\phi_{p'}^{}(\vec k',\vec x)\,d^{3}x+\textrm{H.c.}\right).
\end{align}
\endgroup

First, let us consider the term with $\phi_{lm}^{}(E,\vec x)$ and $\phi_{l'm'}^{}(E',\vec x)$. Using the explicit form of the functions $\phi_{lm}^{}(E,\vec x)$ and $\phi_{l'm'}^{}(E',\vec x)$ (see \eqref{philm} and \eqref{Ylm}), it is not difficult to show that
\begin{equation}
\int\limits_{r>r_{0}}\sqrt{-g}g^{00}\phi_{lm}^{}(E,\vec x)\phi_{l'm'}^{}(E',\vec x)\,d^{3}x=\delta_{ll'}\delta_{m,-m'}\delta(E-E').
\end{equation}
Due to the presence of $\delta(E-E')$ in the latter relation, we have
\begin{equation}
\left(\frac{E\sqrt{E}}{\sqrt{E'}}-\sqrt{EE'}\right)\delta(E-E')=(E-E)\delta(E-E')=0,
\end{equation}
which means that the whole term with the coefficient $\frac{E\sqrt{E}}{\sqrt{E'}}-\sqrt{EE'}$ in \eqref{Hamaux1} vanishes.

Second, let us consider the term with $\phi_{p}^{}(\vec k,\vec x)$ and $\phi_{p'}^{}(\vec k',\vec x)$. It is not difficult to show that (see Appendix~A for analogous calculations)
\begin{equation}\label{orthcondaux2}
\int\limits_{r>r_{0}}\sqrt{-g}g^{00}
\phi_{p}^{}(\vec k,\vec x)\phi_{p'}^{}(\vec k',\vec x)\,d^{3}x=\frac{1}{4\pi k^{2}}\sum\limits_{l=0}^{\infty}(2l+1)e^{i(\pi l+2\tilde\delta_{lp}(k))}P_{l}(\cos\alpha)\delta_{pp'}\delta(k-k'),
\end{equation}
where $\alpha$ is the angle between the vectors $\vec k$ and $\vec k'$ (defined in spherical coordinates in the momentum space by $k$, $\theta_{k}$, $\varphi_{k}$ and $k'$, $\theta_{k}'$, $\varphi_{k}'$ respectively), which looks like \cite{Korn-Korn}
\begin{equation}
\cos\alpha=\cos\theta_{k}\cos\theta_{k}'+\sin\theta_{k}\sin\theta_{k}'\cos(\varphi_{k}-\varphi_{k}').
\end{equation}
Due to the presence of $\delta(k-k')$ in
\eqref{orthcondaux2}, we have
\begin{equation}
\left(\frac{(k^{2}+M^{2})^{\frac{3}{4}}}{({k'}^{2}+M^{2})^{\frac{1}{4}}}
-(k^{2}+M^{2})^{\frac{1}{4}}({k'}^{2}+M^{2})^{\frac{1}{4}}\right)\delta(k-k')=\left(\sqrt{k^{2}+M^{2}}-\sqrt{k^{2}+M^{2}}\right)\delta(k-k')=0,
\end{equation}
which means that the whole term with the coefficient $\frac{(k^{2}+M^{2})^{\frac{3}{4}}}{({k'}^{2}+M^{2})^{\frac{1}{4}}}
-(k^{2}+M^{2})^{\frac{1}{4}}({k'}^{2}+M^{2})^{\frac{1}{4}}$ in \eqref{Hamaux1} also vanishes. Thus, for \eqref{Hamaux1} we get
\begin{align}\nonumber
H&=\frac{1}{2}\sum\limits_{l=0}^{\infty}\sum\limits_{m=-l}^{l}\int\limits_{0}^{M}E\left(a_{lm}^{\dagger}(E)a_{lm}^{}(E)+a_{lm}^{}(E)a_{lm}^{\dagger}(E)\right)dE
\\\label{Hamaux2}
&+\frac{1}{2}\sum\limits_{p=1}^{2}\int\sqrt{k^{2}+M^{2}}
\left(a_{p}^{\dagger}(\vec k)a_{p}^{}(\vec k)+a_{p}^{}(\vec k)a_{p}^{\dagger}(\vec k)\right)d^{3}k.
\end{align}
Passing from $a_{lm}^{}(E)a_{lm}^{\dagger}(E)$ to
$a_{lm}^{\dagger}(E)a_{lm}^{}(E)$ and from $a_{p}^{}(\vec
k)a_{p}^{\dagger}(\vec k)$ to $a_{p}^{\dagger}(\vec
k)a_{p}^{}(\vec k)$ in \eqref{Hamaux2}, and dropping
the irrelevant $c$-number terms, finally we obtain
\begin{equation}\label{Hamiltresult}
H=\sum\limits_{l=0}^{\infty}\sum\limits_{m=-l}^{l}\int\limits_{0}^{M}E\,a_{lm}^{\dagger}(E)a_{lm}^{}(E)\,dE
+\int\sqrt{k^{2}+M^{2}}
\left(a_{1}^{\dagger}(\vec k)a_{1}^{}(\vec k)+a_{2}^{\dagger}(\vec k)a_{2}^{}(\vec k)\right)d^{3}k.
\end{equation}
We see that the resulting Hamiltonian has the standard form.
However, this Hamiltonian implies that there is a degeneracy of
states that are parameterized by the same $\vec
k$. Of course, these states differ in the functions $A_{1}(\vec k,\vec n,r)$ and $A_{2}(\vec k,\vec n,r)$ in
\eqref{scatstates}. Meanwhile, for an observer at large distances
from the black hole, where the terms with the functions $A_{p}(\vec k,\vec n,r)$ can be neglected because of the factor $\frac{1}{r}$,
these different states look just as identical slightly modified plane waves with the
same asymptotic momentum $\vec k$. The origin of this peculiarity will be
briefly discussed in the next section.

\subsection{Brief comparison with the case of Minkowski spacetime}
Let us consider a free real massive scalar field
in Minkowski spacetime. Formally, we can start the
analysis from the solutions in spherical coordinates.
In particular, the properly normalized solutions of the
corresponding radial equation have the form \cite{LL-QM}
\begin{equation}
R_{l}(k,r)=\sqrt{\frac{2}{\pi}}\,(-1)^{l}\left(\frac{r}{k}\right)^{l}\left(\frac{1}{r}\frac{d}{dr}\right)^{l}\frac{\sin kr}{r}
\end{equation}
such that
\begin{equation}
\int\limits_{0}^{\infty}R_{l}(k,r)R_{l}(k',r)\,r^{2}dr=\delta(k-k').
\end{equation}
Using these solutions, we can also build the ``scattering states'' \cite{LL-QM}
\begin{equation}
\frac{1}{(2\pi)^{\frac{3}{2}}}\,e^{i\vec k\vec x}=\frac{1}{4\pi k}\sum\limits_{l=0}^{\infty}(2l+1)\,e^{i\frac{\pi l}{2}}P_{l}\left(\frac{\vec k\vec x}{kr}\right)R_{l}(k,r),
\end{equation}
which are just the exact plane waves. However,
because there exits only {\em one} real radial solution for fixed
$k$ and $l$ (in particular, for $l\neq 0$ this is the consequence of the existence of the
dominant term $\frac{l(l+1)}{r^{2}}$ at $r\to 0$ in the
corresponding equation), for a fixed $\vec k$ it is possible to
build only one state that behaves as a plane wave for
$r\to\infty$ (of course, in this particular case the state behaves
as a plane wave everywhere in space). Thus, for the Hamiltonian we
get the well-known exact result
\begin{equation}
H=\int\sqrt{k^{2}+M^{2}}\,a^{\dagger}(\vec k)a^{}(\vec k)\,d^{3}k
\end{equation}
without any degeneracy of states. Of course, the same reasoning
is valid in the cases of standard short-range spherically symmetric
potentials in Minkowski spacetime, the only difference with the
case of the free field being a nonzero scattering
amplitude. For the standard Coulomb potential, we can also build one scatteringlike state of form \eqref{scatstates} (note that in the standard approach the solution corresponding to the Coulomb potential is considered in a different form \cite{LL-QM}). In contrast to this, in the case of the Schwarzschild
metric there exist {\em two} real radial solution for fixed $k$
and $l$, leading to two scatteringlike states with the same $\vec k$.
It is a purely topological effect, which cannot be eliminated by
moving away from the black hole.

\section{Discussion and conclusion}
In the present paper, we have performed the procedure
of canonical quantization of a real massive scalar field outside
the horizon of an ideal Schwarzschild black hole. We
have shown that the resulting theory turns out to be complete and
self-consistent, i.e., the canonical commutation relations are
satisfied exactly and the Hamiltonian has the standard form
without any peculiarities. Although the scalar field theory we are
working with is local in the sense that it is supposed to contain
only local interactions, it relies on the existence of
solutions of the corresponding equation of motion in
the whole space. The latter means that in this sense the theory is
``global''. On the other hand, though the time
coordinate $t$ can be considered as proper time
only at $r\to\infty$, it does not lead to any contradiction in
performing the canonical quantization procedure. Moreover, since
the initial theory is invariant under the
translations in time $t$, it gives rise to
Hamiltonian \eqref{Hamiltscalar} that is conserved over time,
which is essential for obtaining correct quantum
field theory. Thus, time $t$ can be considered as a global time in
the resulting quantum theory.

It is clear that the results presented in this paper can be easily
reduced to the massless case by setting $M=0$. In this case,
Eq.~\eqref{eqSchr} with \eqref{VSchr1} takes the form of the
Regge-Wheeler equation \cite{RW}. Of course, if $M=0$, then
the ``localized'' states (the first term with the
integral with respect to $E$ in \eqref{Hamiltresult}) are absent in the
theory but all the other conclusions concerning the
properties of the spectrum remain the same.

A feature of the resulting quantum theory is
that the Schwarzschild black hole interior is not necessary for
obtaining correct quantum field theory outside the
black hole and does not affect it. In this sense this picture is
similar to the one discussed in \cite{tHooft3}, the difference being
that we do not need the white hole exterior as well.

Another feature is that in the vicinity of the horizon the scalar
particles become effectively massless, which is a consequence of
the fact that effective potential $V_{l}(z)$ \eqref{VSchr1} in the
radial equation \eqref{eqSchr} is such that $V_{l}(z)\to 0$ as
$z\to-\infty$ (i.e., as $r\to r_{0}$). It means that,
presumably, bound states would decay as they approach the event horizon.

And one more feature, which is even more important,
is connected with the spectrum of quantum states. This spectrum
consists of two branches. The first
branch is the continuous spectrum of states with energies less
than the mass of the field, these states describe
particles that are bound in the vicinity of the horizon. The second
branch is the continuous spectrum of states with
energies larger than the mass of the field. At large
distances from the Schwarzschild black hole, the latter states
look similar to those of the usual particles with definite momenta.
However, it turns out that for a fixed vector $\vec k$ there exist {\em
two} different states corresponding to such particles. Naively one
would expect that if an observer is located far away from the
Schwarzschild black hole, the effects caused by the black hole can
be neglected. Nevertheless, it is not so --- as it has been demonstrated above,
because of the different topological structures of the
Schwarzschild spacetime and Minkowski spacetime (the former is
topologically $R^2\times S^2$, whereas the latter is $R^4 $), the
structures of the spectra in both spacetimes are completely
different, which manifests itself in the additional
degeneracy in the case of the Schwarzschild spacetime. This
effect is a direct consequence of the fact that, as was mentioned
above, the field theory is ``global'' in the sense that in order
to have a consistent classical or quantum effective theory, it is
necessary to have solutions of the corresponding equations of
motion in the whole space. Of course, this conclusion was
rigorously proven only for the simplest case of a real massive
scalar field but we expect
that such a degeneracy of states is inherent to fields of
different types, i.e., to vector and spinor fields too. Moreover,
it is quite possible that the existence of several black holes may
lead to additional degeneracy of quantum states. We expect that
the cross sections of various processes calculated in
such a theory would differ considerably from those in the Standard
Model, even in the case of a single black hole.

We would also like to note that, according to
\cite{Akhmedov:2016uha}, a very compact object differs from a
collapsing star in the existence of discrete energy
levels in the former case. The same situation with the
Schwarzschild black hole
--- the spectrum of ``localized'' states with energies less than
the scalar field mass is also continuous. However, as was
demonstrated above, there exists a more serious difference --- we
do not expect a degeneracy of states in the quantum theory for a
very compact object (at least, for standard compact objects consisting of ordinary matter), because such objects do not change the
spacetime topology. Meanwhile, black holes change the spacetime
topology in such a way that there emerge extra degrees of freedom of the scalar field and, consequently, even in the simplest case of the Schwarzschild black hole there emerges a degeneracy which leads to the consequences discussed above. In principle, the existence of such a degeneracy
poses a question about the existence of black holes with horizons leading to substantial changes in the effective theory even far away from the black hole.

It should be noted that there may exist objects
that have no horizons but also change the spacetime
topology in the same manner, for example, traversable wormholes of
the Morris-Thorne type
\cite{Ellis:1973yv,Bronnikov:1973fh,Morris:1988cz}.\footnote{We
are very grateful to the anonymous referee for indicating this point and suggesting the
Morris-Thorne wormhole as an example.} In the first case,
where the wormhole connects two different universes
(topologically the corresponding spacetime is also $R^2\times
S^2$), the degeneracy of states completely analogous
to the one of the Schwarzschild black hole case is expected, and
the only substantial difference with the latter case is the form
of the effective potential in the corresponding
equation analogous to Eq.~\eqref{eqSchr}. Namely,
for such a wormhole the radial coordinate can be chosen in such a way that
$r\in(-\infty,\infty)$, the corresponding potential $V_{l}(r)$ is
symmetric and has the properties that $V_{l}(r)\to\mu^{2}$ for
$r\to\pm\infty$ and there is a well in the vicinity of $r=0$.
Thus, the doubling of physical states is expected for observers
located far away from the wormhole in both universes.

If the wormhole connects two regions of our
universe, then the situations turns out to be more
involved. The topology of such a spacetime is that
of a noncompact handlebody and clearly differs from the one of
Minkowski spacetime. However, since the wormhole has a finite
volume, whereas different asymptotic regions of space are not
connected only through the wormhole, it is possible that a
degeneracy of states is absent in this case and the wormhole
just imposes additional restrictions on the wave functions of states. In other words, the Schwarzschild black hole
and the wormhole that connects different universes provide new
asymptotic regions in comparison with those in Minkowski
spacetime (i.e., they change the global topology of spacetime),
whereas it is not so for the wormhole that connects two regions of
our universe (i.e., it changes the spacetime
topology only in a finite region of space; this can be easily
understood if one imagines that two exits of the wormhole are
located close to each other). Thus, at the moment we
have no definite answer to the question, whether
there is a degeneracy of states for such a wormhole or not. This
problem, as well as the effects produced by the degeneracy and
possible ways to avoid it, call for a further analysis.

\subsection*{Acknowledgments}
The authors are grateful to S.O. Alexeyev, E.E. Boos, Yu.V. Grats, S.I. Keizerov, V.P.
Neznamov, S.A. Paston, Yu.V. Popov, E.R. Rakhmetov, and E.V. Troitsky for valuable
discussions and comments. The research was carried out within the framework of
the scientific program of the National Center for Physics and
Mathematics, the project ``Particle Physics and Cosmology''.

\section*{Appendix~A: Orthogonality condition for the scatteringlike states}
Let us consider \eqref{orthscatt2} with $\phi_{p}(\vec k,\vec x)$ defined by \eqref{scatstatesdec}. To prove this orthogonality condition, we will go along the lines of the method used in \cite{LL-QM} for a similar proof. Let $\beta$ be the angle between the vectors $\vec k$ and $\vec x$, $\beta'$ is the angle between the vectors $\vec k'$ and $\vec x$, $\alpha$ is the angle between the vectors $\vec k$ and $\vec k'$, and $\tilde\varphi$ is the angle between the planes $(\vec x,\vec k)$ and $(\vec k,\vec k')$. One can check that in such a case the relation
\begin{equation}
\cos\beta'=\cos\beta\cos\alpha+\sin\beta\sin\alpha\cos\tilde\varphi
\end{equation}
between the angles holds, as well as the addition theorem for the Legendre polynomials \cite{Korn-Korn}
\begin{equation}\label{addtheorem}
P_{l'}(\cos\beta')=P_{l'}(\cos\beta)P_{l'}(\cos\alpha)
+2\sum\limits_{m=1}^{l'}\frac{(l'-m)!}{(l'+m)!}P_{l'}^{m}(\cos\beta)P_{l'}^{m}(\cos\alpha)\cos(m\tilde\varphi).
\end{equation}
Using these angles, the integral in the lhs of
\eqref{orthscatt2} can be rewritten as\footnote{One can easily
check that the integrations with respect to $\beta$
and $\tilde\varphi$ in \eqref{orthscatt2app} go over
the total solid angle of the vector $\vec x$.}
\begin{align}\nonumber
\frac{1}{16\pi^{2}kk'}\frac{\sqrt{k}}{\left(k^2+M^{2}\right)^{\frac{1}{4}}}\frac{\sqrt{k'}}{\left({k'}^2+M^{2}\right)^{\frac{1}{4}}}\sum\limits_{l=0}^{\infty}
\sum\limits_{l'=0}^{\infty}(2l+1)(2l'+1)e^{i\left(\frac{\pi(l-l')}{2}+\tilde\delta_{lp}(k)-\tilde\delta_{l'p'}(k')\right)}
\\\label{orthscatt2app}\times
\int\limits_{0}^{2\pi}d\tilde\varphi\int\limits_{0}^{\pi}\sin\beta\,d\beta\int\limits_{r_{0}}^{\infty}dr\frac{r^{3}}{r-r_{0}}
P_{l}(\cos\beta)P_{l'}(\cos\beta')f_{lp}\left(\sqrt{k^{2}+M^{2}},r\right)f_{l'p'}\left(\sqrt{{k'}^{2}+M^{2}},r\right).
\end{align}
Let us take only the angular part of the latter integral, which reads
\begin{equation}
\int\limits_{0}^{2\pi}d\tilde\varphi\int\limits_{0}^{\pi}\sin\beta\,d\beta P_{l}(\cos\beta)P_{l'}(\cos\beta'),
\end{equation}
and substitute \eqref{addtheorem} into it. We get
\begin{align}\nonumber
\int\limits_{0}^{2\pi}d\tilde\varphi\int\limits_{0}^{\pi}\sin\beta\,d\beta P_{l}(\cos\beta)P_{l'}(\cos\beta')&=
2\pi\int\limits_{0}^{\pi}\sin\beta\,d\beta P_{l}(\cos\beta)P_{l'}(\cos\beta)P_{l'}(\cos\alpha)\\\label{angpartortho}&=\frac{4\pi}{2l+1}P_{l}(\cos\alpha)\delta_{ll'}.
\end{align}
Substituting \eqref{angpartortho} into \eqref{orthscatt2app}, we get
\begin{align}\nonumber
\sum\limits_{l=0}^{\infty}\frac{2l+1}{4\pi kk'}\frac{\sqrt{k}}{\left(k^2+M^{2}\right)^{\frac{1}{4}}}\frac{\sqrt{k'}}{\left({k'}^2+M^{2}\right)^{\frac{1}{4}}}\,P_{l}(\cos\alpha)e^{i\left(\tilde\delta_{lp}(k)-\tilde\delta_{lp'}(k')\right)}
\\\label{orthscatt2app2}\times
\int\limits_{r_{0}}^{\infty}dr\frac{r^{3}}{r-r_{0}}f_{lp}\left(\sqrt{k^{2}+M^{2}},r\right)f_{lp'}\left(\sqrt{{k'}^{2}+M^{2}},r\right).
\end{align}
With the orthogonality condition \eqref{orthrad30}, formula \eqref{orthscatt2app2} takes the form
\begin{align}\nonumber
\delta_{pp'}\frac{k}{\sqrt{k^2+M^{2}}}\,\delta\left(\sqrt{k^2+M^{2}}-\sqrt{{k'}^2+M^{2}}\right)\sum\limits_{l=0}^{\infty}\frac{2l+1}{4\pi k^{2}}P_{l}(\cos\alpha)\\\label{orthscatt2app3}=
\delta_{pp'}\delta(k-k')\sum\limits_{l=0}^{\infty}\frac{2l+1}{4\pi k^{2}}P_{l}(\cos\alpha).
\end{align}
Recall that \cite{LL-QM}
\begin{equation}\label{Pdelta}
\frac{1}{4}\sum\limits_{l=0}^{\infty}(2l+1)P_{l}(\cos\alpha)=\delta(1-\cos\alpha).
\end{equation}
With the latter relation, formula \eqref{orthscatt2app3} takes the form
\begin{equation}\label{orthscatt2app4}
\delta_{pp'}\frac{1}{\pi k^{2}}\,\delta(k-k')\delta(1-\cos\alpha).
\end{equation}
It is clear that the delta functions in
\eqref{orthscatt2app4} select exactly $\vec k=\vec k'$, i.e.,
$\frac{1}{\pi k^{2}}\,\delta(k-k')\delta(1-\cos\alpha)$ should
correspond to $\delta^{(3)}(\vec k-\vec k')$ in the initial
variables. It is indeed so, because \cite{LL-QM}
\begin{align}\nonumber
\int d^{3}k
\left(\frac{1}{\pi k^{2}}\,\delta(k-k')\delta(1-\cos\alpha)\right)&
\\\label{orthscatt2app5}
=2\pi\int\limits_{0}^{\infty}k^{2}dk\int\limits_{0}^{\pi}\sin\alpha\,d\alpha\left(\frac{1}{\pi k^{2}}\,\delta(k-k')\delta(1-\cos\alpha)\right)&=2\int\limits_{-1}^{1}\delta(1-y)dy=1,
\end{align}
where the spherical coordinate system is chosen such that its ``$z$-axis'' coincides with the vector $\vec k'$ and the prescription $\int_{-1}^{1}\delta(1-y)dy=1/2$ is used. Relation \eqref{orthscatt2app5} finalizes the proof of \eqref{orthscatt2}.

\section*{Appendix~B: Completeness relation involving the scatteringlike states}
First, let us consider the second part in the lhs of \eqref{complete4}
\begin{equation}\label{complete4app1}
\sum\limits_{p=1}^{2}\int\phi_{p}^{*}(\vec k,\vec x)\phi_{p}^{}(\vec k,\vec y)\,d^{3}k.
\end{equation}
Let $\beta$ be the angle between the vectors $\vec x$ and $\vec k$, $\beta'$ is the angle between the vectors $\vec y$ and $\vec k$, $\gamma$ is the angle between the vectors $\vec x$ and $\vec y$, and $\tilde\varphi$ is the angle between the planes $(\vec k,\vec x)$ and $(\vec x,\vec y)$. One can check that in such a case the relation
\begin{equation}
\cos\beta'=\cos\beta\cos\gamma+\sin\beta\sin\gamma\cos\tilde\varphi
\end{equation}
between the angles holds. Using these angles (contrary to the case discussed in Appendix~A, here the angles $\beta$, $\beta'$ and $\tilde\varphi$ parameterize the momentum space), the integral \eqref{complete4app1} can be rewritten as
\begin{align}\nonumber
\frac{1}{16\pi^{2}}\sum\limits_{p=1}^{2}\sum\limits_{l=0}^{\infty}\sum\limits_{l'=0}^{\infty}(2l+1)(2l'+1)e^{i\frac{\pi(l-l')}{2}}
\int\limits_{0}^{2\pi}d\tilde\varphi\int\limits_{0}^{\pi}\sin\beta\,d\beta P_{l}(\cos\beta)P_{l'}(\cos\beta')\\\label{complete4app2}\times\int\limits_{0}^{\infty}dk\,e^{i\left(\tilde\delta_{lp}(k)-\tilde\delta_{l'p}(k)\right)}
\frac{k}{\sqrt{k^2+M^{2}}}
f_{lp}\left(\sqrt{k^{2}+M^{2}},r\right)f_{l'p}\left(\sqrt{k^{2}+M^{2}},r'\right),
\end{align}
where $r=|\vec x|$, $r'=|\vec y|$. Using the addition theorem for the Legendre polynomials \cite{Korn-Korn}
\begin{equation}\label{addtheorem2}
P_{l'}(\cos\beta')=P_{l'}(\cos\beta)P_{l'}(\cos\gamma)
+2\sum\limits_{m=1}^{l'}\frac{(l'-m)!}{(l'+m)!}P_{l'}^{m}(\cos\beta)P_{l'}^{m}(\cos\gamma)\cos(m\tilde\varphi)
\end{equation}
and \eqref{angpartortho}, formula \eqref{complete4app2} can be brought to the form
\begin{align}\nonumber
\frac{1}{4\pi}\sum\limits_{p=1}^{2}\sum\limits_{l=0}^{\infty}(2l+1)
P_{l}(\cos\gamma)\int\limits_{0}^{\infty}dk\,\frac{k}{\sqrt{k^2+M^{2}}}
f_{lp}\left(\sqrt{k^{2}+M^{2}},r\right)f_{lp}\left(\sqrt{k^{2}+M^{2}},r'\right)\\\label{complete4app3}
=\frac{1}{4\pi}\sum\limits_{l=0}^{\infty}(2l+1)
P_{l}(\cos\gamma)\sum\limits_{p=1}^{2}\int\limits_{M}^{\infty}
f_{lp}\left(E,r\right)f_{lp}\left(E,r'\right)dE.
\end{align}

Now let us consider the first part in the lhs of \eqref{complete4}
\begin{equation}\label{complete4app4}
\sum\limits_{l=0}^{\infty}\sum\limits_{m=-l}^{l}\int\limits_{0}^{M}\phi_{lm}^{*}(E,\vec x)\phi_{lm}^{}(E,\vec y)dE.
\end{equation}
Using the explicit form of spherical harmonics \eqref{Ylm}, we can get
\begin{align}\nonumber
&\sum\limits_{m=-l}^{l}Y_{lm}^{*}(\theta,\varphi)Y_{lm}(\theta',\varphi')\\\label{complete4app5}
&=\frac{2l+1}{4\pi}\left(P_{l}(\cos\theta)P_{l}(\cos\theta')+2\sum\limits_{m=1}^{l}
\frac{(l-m)!}{(l+m)!}\,P_{l}^{m}\left(\cos\theta\right)P_{l}^{m}\left(\cos\theta'\right)\cos(m(\varphi-\varphi'))\right).
\end{align}
It is known that for the standard spherical coordinates the relation
\begin{equation}\label{anglesappBrel}
\cos\gamma=\cos\theta\cos\theta'+\sin\theta\sin\theta'\cos(\varphi-\varphi')
\end{equation}
holds, where $\gamma$ is the angle between the vectors $\vec x$
and $\vec y$ defined by $r$, $\theta$, $\varphi$, and $r'$,
$\theta'$, $\varphi'$, respectively \cite{Korn-Korn}. So, according
to relation \eqref{anglesappBrel}, the addition theorem for the
Legendre polynomials takes the form
\begin{equation}\label{addtheorem3}
P_{l}(\cos\gamma)=P_{l}(\cos\theta)P_{l}(\cos\theta')
+2\sum\limits_{m=1}^{l}\frac{(l-m)!}{(l+m)!}P_{l}^{m}(\cos\theta)P_{l}^{m}(\cos\theta')\cos(m(\varphi-\varphi')).
\end{equation}
With \eqref{addtheorem3}, formula \eqref{complete4app5} can be rewritten as
\begin{equation}\label{complete4app6}
\sum\limits_{m=-l}^{l}Y_{lm}^{*}(\theta,\varphi)Y_{lm}(\theta',\varphi')=\frac{2l+1}{4\pi}P_{l}(\cos\gamma).
\end{equation}
Thus, for \eqref{complete4app4} we obtain
\begin{equation}\label{complete4app7}
\sum\limits_{l=0}^{\infty}\sum\limits_{m=-l}^{l}\int\limits_{0}^{M}\phi_{lm}^{*}(E,\vec x)\phi_{lm}^{}(E,\vec y)dE=\frac{1}{4\pi}\sum\limits_{l=0}^{\infty}(2l+1)P_{l}(\cos\gamma)\int\limits_{0}^{M}f_{l}(E,r)f_{l}(E,r')dE.
\end{equation}

Combining \eqref{complete4app3} and \eqref{complete4app7}, for the lhs of \eqref{complete4} we get
\begin{equation}\label{complete4app8}
\frac{1}{4\pi}\sum\limits_{l=0}^{\infty}(2l+1)P_{l}(\cos\gamma)\left(\int\limits_{0}^{M}f_{l}(E,r)f_{l}(E,r')dE
+\sum\limits_{p=1}^{2}\int\limits_{M}^{\infty}f_{lp}\left(E,r\right)f_{lp}\left(E,r'\right)dE\right).
\end{equation}
With the completeness relation \eqref{complete2} and with \eqref{Pdelta}, formula \eqref{complete4app8} can be brought to the form
\begin{equation}\label{complete4app9}
\frac{1}{4\pi}\sum\limits_{l=0}^{\infty}(2l+1)P_{l}(\cos\gamma)\frac{r-r_{0}}{r^{3}}\,\delta(r-r')=
\frac{1}{\pi}\,\delta(1-\cos\gamma)\frac{r-r_{0}}{r^{3}}\,\delta(r-r').
\end{equation}
Like in the case of \eqref{orthscatt2app4}, the delta
functions in \eqref{complete4app9} select exactly $\vec x=\vec
y$. The calculation (compare with \eqref{orthscatt2app5})
\begin{align}\nonumber
\int\limits_{r>r_{0}}d^{3}x\sqrt{-g}\,g^{00}\left(\frac{1}{\pi}\,\delta(1-\cos\gamma)\frac{r-r_{0}}{r^{3}}\,\delta(r-r')\right)&\\\label{complete4app10}=
2\pi\int\limits_{r_{0}}^{\infty}dr\frac{r^{3}}{r-r_{0}}\int\limits_{0}^{\pi}d\gamma\sin\gamma
\left(\frac{1}{\pi}\,\delta(1-\cos\gamma)\frac{r-r_{0}}{r^{3}}\,\delta(r-r')\right)&=1,
\end{align}
in which the spherical coordinate system is chosen such that its ``$z$-axis'' coincides with the vector $\vec y$ and again the prescription $\int_{-1}^{1}\delta(1-y)dy=1/2$ is used, finalizes the proof of \eqref{complete4}.

\end{document}